\documentclass[runningheads]{llncs}
\usepackage{hyperref}
\usepackage{graphicx}
\usepackage{xcolor}
\usepackage[nice]{nicefrac}
\usepackage{amsmath}
\usepackage{amsfonts}
\usepackage{mathtools}

\begin{document}
\title{Contextual Attention Network: Transformer Meets U-Net}
%

\author{Reza Azad\inst{1} \and
Moein Heidari\inst{2} \and
Yuli Wu\inst{1} \and Dorit Merhof\inst{1,3}}


\institute{Institute of Imaging and Computer Vision,
RWTH Aachen University, Germany\and
School of Electrical Engineering, Iran University of Science and Technology, Iran, moein\_heidari@elec.iust.ac.ir \and
Fraunhofer Institute for Digital Medicine MEVIS, Bremen, Germany\\\email{\{azad, yuli.wu,dorit.merhof\}@lfb.rwth-aachen.de}}

\maketitle
\begin{abstract}
Currently, convolutional neural networks (CNN) (e.g., U-Net) have become the de facto standard and attained immense success in medical image segmentation. However, as a downside, CNN based methods are a double-edged sword as they fail to build long-range dependencies and global context connections due to the limited receptive field that stems from the intrinsic characteristics of the convolution operation. Hence, recent articles have exploited Transformer variants for medical image segmentation tasks which open up great opportunities due to their innate capability of capturing long-range correlations through the attention mechanism. Although being feasibly designed, most of the cohort studies incur prohibitive performance in capturing local information, thereby resulting in less lucidness of boundary areas. In this paper, we propose a contextual attention network to tackle the aforementioned limitations. The proposed method uses the strength of the Transformer module to model the long-range contextual dependency. Simultaneously, it utilizes the CNN encoder to capture local semantic information. In addition, an object-level representation is included to model the regional interaction map. The extracted hierarchical features are then fed to the contextual
attention module to adaptively recalibrate the representation space using the local information. Then, they emphasize the informative regions while taking into account the long-range contextual dependency derived by the Transformer module. We validate our method on several large-scale public medical image segmentation datasets and achieve state-of-the-art performance. We have provided the implementation code in \href{https://github.com/rezazad68/TMUnet}{\textcolor{red} {github}.}

\end{abstract}

\section{Introduction}
Convolutional neural networks (CNNs), and more specifically, fully convolutional networks, have shown prominence in the majority of medical image segmentation applications. As a variant of these architectures, U-Net~\cite{ronneberger2015u} has rendered notable performance and has been extensively utilized across a wide range of medical domains \cite{valanarasu2020kiu,huang2020unet,cai2020dense,feyjie2020semi,azad2021smu}.
In spite of their superb performance, CNN-based approaches suffer from a limitation in modeling the long-range semantic dependencies due to a confined receptive field size (even with dilated/atrous sampling~\cite{chen2017deeplab}) and due to the nature of the convolution layer. Therefore, such a deficiency in capturing multi-scale information yields a performance degradation in the segmentation of complex structures, with variation in shapes and scales.
Different methods have been proposed to solve the problem of the restricted receptive field of regular CNNs in recent years~\cite{chen2017deeplab,li2021accurate}.
Wang et al.~\cite{wang2018non} extended the self-attention concept into the spatial domain to model non-local properties of images by devising a non-local module that can be easily integrated into
existing network designs. As a result of the prominent performance of the attention mechanism, a line of research has studied bridging the gap between the attention mechanism and CNNs in medical image segmentation~\cite{sinha2020multi,cai2020ma,azad2021deep}.
To overcome the aforementioned limitation of CNNs, recently
proposed Transformer-based architectures that leverage the self-attention mechanism to construct the contextual representations have been utilized. 
Transformers, unlike regular CNNs, are not only capable of modeling global contexts but also of leveraging to the versatile local information. Inspired by this, numerous studies have attempted to adapt Transformers for various image recognition tasks~\cite{dosovitskiy2020image,chen2021crossvit}.
Moreover, Transformer-based models have recently gained growing attention ahead of their CNN counterpart in medical image segmentation~\cite{valanarasu2021medical,chen2021transunet}.
As an example of an alternative perspective by treating semantic segmentation using Transformers, \cite{zheng2021rethinking} proposed to model semantic segmentation as a sequence-to-sequence prediction task. Although the network is well designed to model the global contextual representation, it pays less attention to the local information and is, consequently, less precise in the boundary area. 
Additionally, case studies have been established to investigate the amalgamation of Transformers and U-Net in medical image segmentation~\cite{chen2021transunet,hatamizadeh2022unetr}.
Valanarasu et al.~\cite{valanarasu2021medical} introduce a Local-Global training methodology for Transformers to learn both global and local features, respectively. However, this approach fails to model object-level interaction and renders a poor performance in the case of overlapping objects of interests.

What all these methods have in common is their limitation in designing a specific mechanism to adaptively combine the global and local contextual representations. Particularly, a mechanism to jointly model the local semantic CNN representation along with the global contextual features derived from the Transformer module is critical for a task-specific purpose. To address this limitation, we propose the contextual attention network. 
Our design offers a two-stream pipeline, where in the first stream we utilize a CNN module to extract local semantic information and the object-level interaction map, while the second path incorporates the Transformer module to capture long-range contextual representations. Our Transformer module produces an image-level contextual representation \emph{(ICR)} to construct the spatial dependency map in the image level and it produces regional importance coefficients \emph{(RIC)} to model the importance of each region. In contrast to \cite{valanarasu2021medical} which simply concatenates the local and global features, our method utilizes a contextual attention module to adaptively scale the feature maps and emphasizes on the important regions. Through extensive experiments, our empirical findings validate that our method is able to pay more attention to the overlapped boundary area while providing a strong semantic segmentation map.
Our main contributions are summarized as follows:
\begin{itemize}
\item  A contextual attention mechanism to adaptively aggregate pixel, object and image level features
\item  Coupling Transformer module with the CNN encoder to model object-level interaction
\item  State-of-the-art results on the public datasets along with a publicly-available implementation source code
\end{itemize}

\section{Proposed Method}
\label{method}
\begin{figure}[ht]
	\centering
	\begin{tabular}{cc}
		\includegraphics[width=1\textwidth]{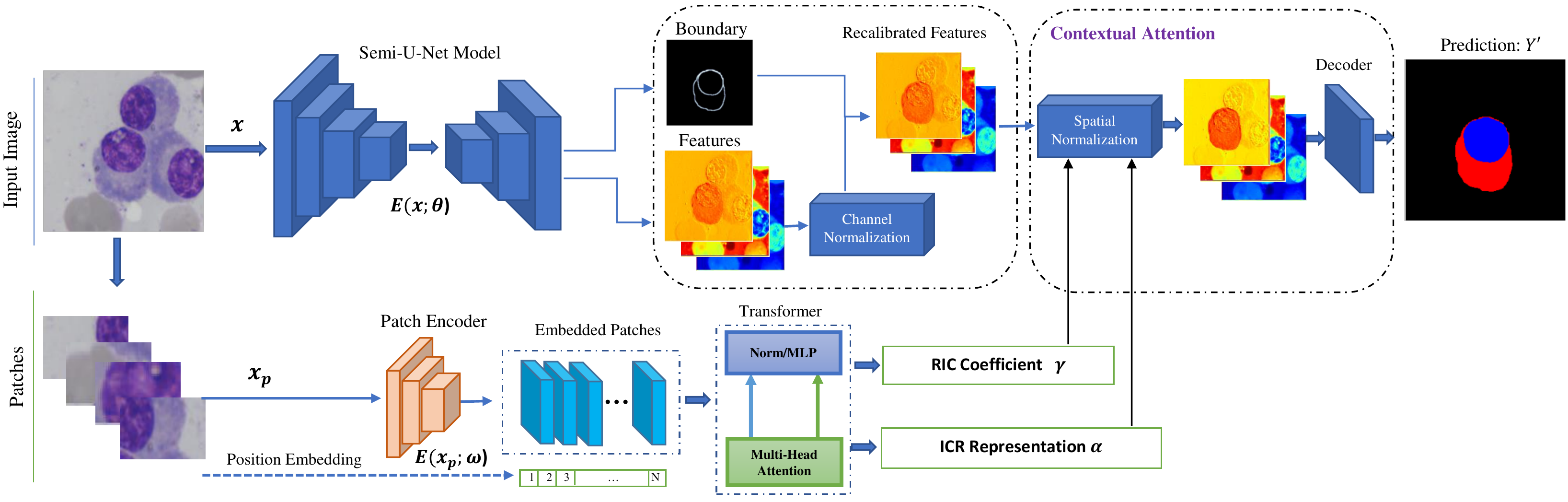}&
	\end{tabular}
	\caption{Illustration of the proposed approach for medical image segmentation with the incorporation of the global contextual representation into local representative features.}
	\label{fig:method}
\end{figure}

The proposed network architecture is depicted in Figure~\ref{fig:method}. Our proposed architecture offers an end-to-end training strategy to adaptively incorporate the global contextual representation into local representative features derived from the CNN module. Our design proposes a contextual attention mechanism for feature recalibration and boundary-aware semantic segmentation. We will discuss each part in the following subsections. 

\subsection{CNN Representation}
As illustrated in Figure~\ref{fig:method}, our proposed method consists of two encoding streams, where in the first path we deploy the semi U-Net structure \cite{ronneberger2015u} to extract CNN representation. Given an image $\mathbf{x} \in \mathcal{R}^{H \times W \times C}$ with spatial dimension $H$ and $W$, and $C$ channels, our CNN encoder $E_{\theta}$ applies a series of convolutional blocks to model the pixel-level contextual representations. The locality nature of the convolutional operation usually limits the strength in modelling the object-level interaction. To include such representations, we model the object-level interaction by learning the boundary heatmap:

\begin{equation}
B = \sigma(Conv_{b}(E(x;\theta ))), B \in \mathcal{R}^{H\times W \times 1}
\end{equation}
where $\sigma$ shows the sigmoid activation and the $Conv_{b}(.)$ shows a $1 \times 1$ kernel convolutional operation. This additional head enables the regional interaction and provides a surrogate signal for modelling the regional contextual dependency.

\subsection{Long-range Contextual Representation}
To learn the long-range contextual dependency we include the Transformer module in the bottom stream of our proposed pipeline. To prepare the input for the Transformer module, we divide the input image $\mathbf{x} \in \mathcal{R}^{H \times W \times C}$ into flattened uniform non-overlapping patches $\mathbf{x}_{p} \in \mathcal{R}^{N \times\left(p^{2} \cdot C\right)}$, where $p \times p$ denotes the dimension of each patch and $N=[\frac{H W}{p^{2}}]$ is the length of image sequence. Afterwards, using a patch encoder $E(x_p; \omega)$, we project the patches into a $K$ dimensional embedding space. 
In order to maintain the spatial information of each patch, we learn a 1-D positional embedding ${I}_{\mathrm{pos}} \in \mathcal{R}^{N \times K}$ 
which is subsequently added to the patch-embedding to preserve positional information ${t}_{0}=\left[{x}_{p}^{1} {I} ; {x}_{p}^{2}{I} ; \cdots ; {x}_{p}^{N} {I}\right]+{I}_{p o s}$, where ${I} \in \mathcal{R}^{\left(p^{2} \cdot C\right) \times K}$ designates the projected patch embedding. 
We then exploit a stack of Transformer blocks encompassing a multi-head self-attention (MSA) and a multilayer perceptron (MLP) to learn the long-range contextual representation. 
An MSA layer is composed of $M$ parallel self-attention heads to scale the embedded patches:
${t}_{i}^{\prime}=\mathrm{MSA}\left(\mathrm{Norm}\left(\mathbf{t}_{i-1}\right)\right)+\mathbf{t}_{i-1}, \quad i=1 \ldots L$. 
Next, the MLP modules learn the long-range contextual dependency by:
${t}_{i}=\mathrm{MLP}\left(\mathrm{Norm}\left({t}_{i}^{\prime}\right)\right)+{t}_{i}^{\prime}, \quad i=1 \ldots L$, where $\mathrm{Norm}()$ denotes layer normalization \cite{ba2016layer}, and ${t}_{i} \in \mathcal{R}^{\frac{H W}{p^{2}} \times d}$ shows the encoded semantic representation in $d$ dimensional space.
In addition to the encoded features, we model the image-level contextual representation \emph{(ICR)} by reshaping \emph{(Re)} the feature and applying a $1 \times 1$ convolutional operation \emph{($Conv_I$)} :

\begin{equation}
 ICR = \sigma(Conv_I{(Re(t_L))}), ICR \in \mathcal{R}^{H \times W \times 1}  
\end{equation}

We use \emph{ICR} to construct the spatial dependency map in the image level to later normalize the feature set generated by the CNN module. We further define the region importance coefficients \emph{(RIC)} to model the distribution of foreground pixels in each region.
The objective of the \emph{RIC} coefficient is to provide a supervisory signal to guide the contextual attention module in determining the important regions (Eq.~\ref{eq:ric}). 
$Conv_R$ shows a $1 \times 1$ convolutional operation.

\begin{equation}\label{eq:ric}
 RIC = \sigma(Conv_R{(t_L)}),\;  RIC \in \mathcal{R}^{ \frac{HW}{p^2} \times 1}  
\end{equation}

\subsection{Contextual Attention Module}
To adaptively aggregate the extracted features, we propose the contextual attention module. The importance of each feature set should be in line with the task at hand, thus, our proposed module utilizes the following two-level normalization steps:
First, it recalibrates the CNN representation for a pixel-level object understanding, then it performs a spatial normalization to selectively emphasize the long-range contextual dependencies inside the feature set. Following the squeeze and excitation~\cite{hu2018squeeze}, we define the channel-wise normalization weights ($w_{ch}$) as:

\begin{equation}
w_{ch}=\sigma\left(\mathbf{W}_{2} \delta\left(\mathbf{W}_{1} G A P{(f)}\right)\right)
\end{equation}
where $GAP$ shows the global average pooling operation applied to the CNN features (f), \textbf{W1} and \textbf{W2} are the learning parameters, and $\delta$ and $\sigma$ are the Sigmoid and ReLU activation functions. We form the normalized features by: 
$ {f'}=w_{ch} \cdot f$. To emphasize the boundary area, we add the boundary representation to the normalized feature, $\tilde{f}=f'+B$. The objective of the boundary feature is to emphasize the boundary regions and guide the model to precisely separate the overlapping objects (e.g. object-level interaction). Next, using the feature set derived from the Transformer module, we perform the spatial normalization. To this end, first we multiply the \emph{RIC} coefficient with the corresponding regions in $\tilde{f}$ to scale the representation based on the regional importance, $f_{sn} = RIC\cdot\tilde{f}$. To further incorporate the long-range dependency, we concatenate the \emph{ICR} representation with the $f_{sn}$ and then apply the convolutional kernel followed by the batch normalization (BN) and activation function to perform a non-linear aggregation:
\begin{equation}
\tilde{f}_{sn} = \delta(BN(Conv(ICR, f_{sn}))
\end{equation}
The $Conv$ is the $1 \times 1$ convolutional operation. The resulting feature set contains both local semantic and global contextual representations which are selectively combined to perform the semantic segmentation task.
Subsequently, we apply the decoder block to the extracted features to predict the segmentation mask,  $Y' = D(\tilde{f}_{sn}; \gamma)$. The joint objective loss function that we optimize during the training is as follows:

\begin{equation}
\mathcal{L}_{\mathrm{joint}} = \lambda_1 \mathcal{L}_{\mathrm{segmentation}}+\lambda_2\mathcal{L}_{\mathrm{boundary}}+\lambda_3\mathcal{L}_{\mathrm{RIC}}
\end{equation}
where $\mathcal{L}_{\mathrm{segmentation}}$ calculates the Cross-entropy loss between the predicted mask and the ground truth, $\mathcal{L}_{\mathrm{boundary}}$ shows the binary cross-entropy loss for the boundary prediction, and $\mathcal{L}_{\mathrm{RIC}}$ calculates the MSE loss between the distribution of foreground pixels in each image patch and the corresponding predicted one. 
We use coefficients $\lambda_{i}, i \in\{1, 2,3\}$ to weight each loss.

\section{Experiments}

\subsection{Dataset}
\label{dataset}
\textbf{Skin Lesion Segmentation}:
Automatic skin lesion segmentation is one of the most demanding tasks in medical image analysis for accurate diagnosis and treatment. In this respect, we focus on three challenge benchmarks: ISIC 2017 \cite{codella2018skin}, ISIC 2018 \cite{codella2019skin} and PH2 \cite{mendoncca2013ph}.
Following the literature work~\cite{asadi2020multi,azad2019bi} for each series, we divide the dataset into train, validation, and test sets accordingly. We use the same setting for a fair evaluation and downsize the original images from the resolution of $576\times767$ pixels to  $256\times256$ pixels in the pre-processing step. 
\\
\textbf{Multiple Myeloma Segmentation}: The proposed method is also evaluated on multiple myeloma cell segmentation grand challenges \cite{gupta2018pcseg}, which are provided by the SegPC 2021 (Segmentation of Multiple Myeloma Plasma Cells in Microscopic Images). Images in this dataset were captured from bone marrow aspirate slides of patients diagnosed with Multiple Myeloma (MM), a type of white blood cell cancer. 
Using the pipeline from \cite{bozorgpour2021multi}, we follow the same strategy as \cite{azad2021deep} and split the original training dataset (290 images) into a training and validation set, and evaluate our method on the original validation set as our new test set.

\subsection{Experimental Set-up}
\textbf{Network Details and Training Process}: 
As depicted in Figure~\ref{fig:method}, our model uses both U-Net and Transformer modules to semantically label the input image. For the U-Net model, we use the Resent encoder~\cite{he2016deep} pre-trained on ImageNet and a four-blocks decoder module to generate the segmentation mask. Simultaneously, the Transformer structure follows the common implementation of the Vision Transformer with M (experimentally 4) heads. The implementation is performed in PyTorch and the results are carried out on a single GPU system with Nvidia RTX 3090. The model is trained end-to-end employing the Adam
optimizer, batch size 4 and a learning rate $10^{-4}$ for 100 epochs.\\
\textbf{Evaluation Protocol}: 
Our evaluation takes into account the evaluation metrics used in the respective challenges, which comprises several well-known segmentation metrics, including 
$\mathrm{sensitivity}=\frac{\mathrm{TP}}{\mathrm{TP}+\mathrm{FN}}$, 
$\mathrm{specificity}=\frac{\mathrm{TN}}{\mathrm{TN}+\mathrm{FP}}$, 
$\mathrm{accuracy}=\frac{\mathrm{TP}+\mathrm{TN}}{\mathrm{TP}+\mathrm{TN}+\mathrm{FP}+\mathrm{FN}}$, $mIOU=\frac{TP}{TP + FP + FN}$ , and $D S C=\frac{2 TP }{2 TP + FP + FN}$ scores, where $TP$ indicates true samples which are correctly classified as true, $TN$ stands for the correct classification of the negative samples, $FP$ and $FN$ show the wrongly classified samples respectively.  

\subsection{Results}

\textbf{Quantitative results}: 
We provide qualitative comparisons on the benchmarks introduced in Section~\ref{dataset}. 
Starting from the skin lesion segmentation scenario, Table~\ref{tab6} reports the comparison results on the three datasets, namely, ISIC 2017, ISIC 2018 and PH2. We exploited different evaluation metrics to accomplish a general and fair comparison.
The baseline approach is simply the U-Net method without any of the proposed modules. Overall, our method attains a superior global performance across all datasets and evaluation metrics and most of the outperforming margins are statistically significant. We also observed that our method outperforms the Transformer~\cite{wu2022fat,chen2021transunet,valanarasu2021medical} counterparts in almost all skin lesion segmentation benchmarks, which further proves the effectiveness of our design compared to other Transformer-based models.
Note that, in contrast to \cite{wu2022fat}, we attained eminent performance without exploiting any augmentation strategy.
Additionally, Table~\ref{tab:segPC} lists the quantitative results of different alternative methods and the proposed network on the SegPC dataset. The results demonstrate that the proposed network attains better results than the other approaches by achieving significant performance gains over the baseline. It is worth mentioning that the SegPC dataset contains samples with high overlaps and the effectiveness of our method is statistically significant in comparison to the SOTA approaches.   


\begin{table}[t] 
	\caption{Performance comparison of the proposed method vs. state-of-the-art methods on skin lesion segmentation benchmarks.}\label{tab6}

	\resizebox{\textwidth}{!}{
		\begin{tabular}{c||c||c||c}
			\hline
			{\begin{tabular}{cccc}
					\multicolumn{4}{c}{\textbf{Articles}} \\
					\hline
					\textbf{} & \textbf{} & \textbf{}&\textbf{} \\
					\hline
				U-Net~\cite{ronneberger2015u} \\
				Att U-Net~\cite{oktay2018attention}\\
				DAGAN~\cite{lei2020skin}\\
				TransUNet~\cite{chen2021transunet}\\
				MCGU-Net~\cite{asadi2020multi}  \\
				MedT~\cite{valanarasu2021medical}\\
				FAT-Net~\cite{wu2022fat}\\
				\hline
				\multicolumn{4}{c}{\textbf{Proposed}} \\
				\hline
	
				\end{tabular}
			} &
			{\begin{tabular}{cccc}
					\multicolumn{4}{c}{\textbf{ISIC 2017}} \\
					\hline
					\textbf{DSC} & \textbf{SE} & \textbf{SP}&\textbf{ACC} \\
					\hline
                     0.8159 & 0.8172 & 0.9680 & 0.9164\\
                     0.8082 & 0.7998 & 0.9776 & 0.9145\\
                     0.8425 & 0.8363 & 0.9716 & 0.9304\\
                     0.8123&0.8263&0.9577&0.9207\\
					 0.8927 & 0.8502 & \textbf{0.9855} &  0.9570\\
					 0.8037 & 0.8064 & 0.9546 & 0.9090\\
					 0.8500 & 0.8392 & 0.9725 & 0.9326\\
					\hline
					\textbf{0.9164} & \textbf{0.9128} & 0.9789 &  \textbf{0.9660}\\
					\hline

				\end{tabular}
			} &
			{\begin{tabular}{cccc}
					\multicolumn{4}{c}{\textbf{ISIC 2018}} \\
					\hline
					\textbf{DSC} & \textbf{SE} & \textbf{SP} & \textbf{ACC} \\
					\hline
		             0.8545 & 0.8800 & 0.9697 &  0.9404  \\
					 0.8566 & 0.8674 & 0.9863 & 0.9376 \\
				     0.8807 & 0.9072 & 0.9588 & 0.9324 \\
			         0.8499 & 0.8578 & 0.9653 & 0.9452\\
				     0.895 & 0.848 & \textbf{0.986} & 0.955 \\
				     0.8389 & 0.8252 & 0.9637 & 0.9358\\
				     0.8903 & \textbf{0.9100} & 0.9699 & 0.9578\\
				    \hline
				    \textbf{0.9059} & 0.9038 & 0.9746 & \textbf{0.9603}\\
				   \hline

				\end{tabular}
			} &
			{\begin{tabular}{cccc}
					\multicolumn{4}{c}{\textbf{PH2}} \\
					\hline
					\textbf{DSC} & \textbf{SE} & \textbf{SP} & \textbf{ACC} \\
					\hline
		             0.8936 & 0.9125 & 0.9588 & 0.9233\\
		             0.9003 & 0.9205 & 0.9640 & 0.9276\\
	                 0.9201&0.8320&0.9640&0.9425\\
	                 0.8840&0.9063&0.9427&0.9200\\					 
					 0.9263 & 0.8322 & 0.9714  & 0.9537\\
					 0.9122 & 0.8472 & 0.9657  & 0.9416\\
					 \textbf{0.9440} & \textbf{0.9441} & 0.9741 & \textbf{0.9703}\\
					\hline
					0.9414 & 0.9395 & \textbf{0.9756} & 0.9647\\
					\hline
				\end{tabular}
			} \\
		\end{tabular}
		}
	\end{table}

\begin{figure}[ht]
	\centering
	\begin{tabular}{cc}
		\includegraphics[width=1\textwidth]{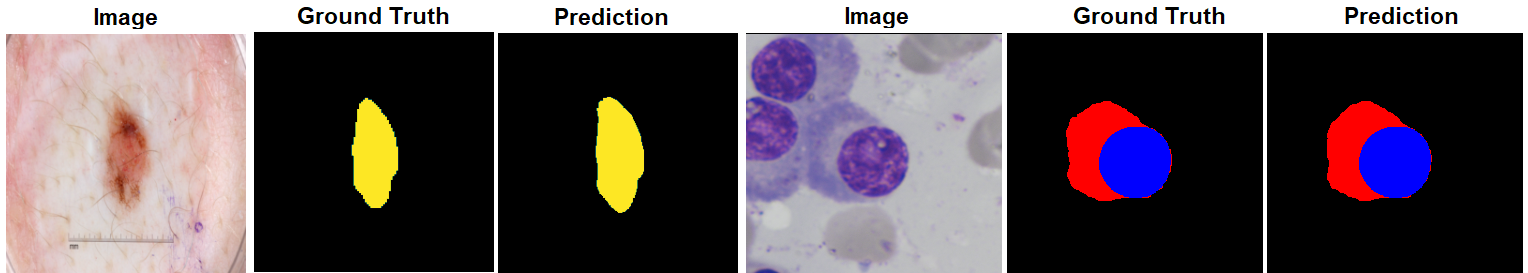}&
	\end{tabular}
	\caption{Prediction results of the proposed method on both skin lesion segmentation and multiple Myeloma instance segmentation.}
	\label{fig:predict}
\end{figure}

\begin{table}
\centering
    \vspace*{-\baselineskip}
	\caption{Performance evaluation on the SegPC challenge (best result is highlighted).}
	\begin{tabular}{cc}
		\hline
		\textbf{Methods} & \textbf{mIOU}\\
		\hline
		Frequency recalibration U-Net \cite{azad2021deep} &0.9392\\		
		XLAB Insights  \cite{bozorgpour2021multi} & 0.9360 \\
		DSC-IITISM  \cite{bozorgpour2021multi} &0.9356\\
		Multi-scale attention deeplabv3+ \cite{bozorgpour2021multi} &0.9065\\
		U-Net \cite{ronneberger2015u} &0.7665\\
		\textbf{Baseline} & 0.9172 \\
		\hline
		\textbf{Proposed}& \textbf{0.9395}\\
		\hline
	\end{tabular}
	\label{tab:segPC}
\end{table}
\textbf{Qualitative results}:
Visual segmentation results on both tasks are illustrated in Figure~\ref{fig:predict}. Clearly, our proposed model produces smooth segmentation results for both tasks and performs well in the boundary area for separating the object of interest from the background. This fact reveals the importance of both Transformer modules for long-range contextual dependency (encouraging object learning) and the combination of boundary modules with the convolutional features maps in precise boundary recovery. Specifically, for the SegPC dataset, we observed that the proposed method segments the myeloma instances from a highly overlapped background with high precision. During our experimental visualization, we also observed that, compared to the U-Net model, the proposed structure produces robust segmentation results even with a noisy annotation, which is a common scenario in the medical domain. Comparable to recent work~\cite{valanarasu2021medical,chen2021transunet}, our results also suggest that coupling the Transformer module with the CNN segmentation model can provide an additional input signal for a reliable and robust segmentation architecture.

\subsection{Ablation Study}
The suggested network comprises the Transformer, the boundary and the attention add-on modules that are included for learning both local and global feature sets. To experimentally evaluate the effect and contribution of each module in the generalization performance, we selectively remove any of the modules, as shown in Table~\ref{table:2}. The qualitative finding suggests that removing any of the modules from the architecture results in a performance loss. More precisely, it can be observed that removing the Transformer module largely decreases the importance of the global attention maps and results in a clear performance drop. However, by including the Transformer module the network takes into account the strength of the global attention mechanism in combination with the local effectiveness of the convolution feature maps in learning generic and rich feature sets for precise localization ability. It is also worthwhile to mention that the combination of the proposed modules decreases the number of wrong predictions and the number of isolated FP (cf. Table~\ref{tab6}) due to the strength of long-range contextual dependency maps. 

\begin{table*}[h!]
\caption{Effect of eliminating each module on the overall performance of the proposed method. We report the results on ISIC 2018 dataset.}
\resizebox{\textwidth}{!}
{
    \setlength{\tabcolsep}{10pt}
    \label{table:2}
    \scalebox{0.25}
    {
        \begin{tabular}{lcccccccc}
        & \multicolumn{3}{c}{Module}
        \\
        \hline
        & Boundary & Transformer & Contextual Attention & DSC
        \\
        \hline 
        & $\textcolor{red}{(\times)}$ &  $\textcolor{green}{(\sqrt)}$ & $\textcolor{green}{(\sqrt)}$ & 0.905 
        \\
        & $\textcolor{green}{(\sqrt)}$ & $\textcolor{red}{(\times)}$ & $\textcolor{green}{(\sqrt)}$ & 0.896
        \\
        & $\textcolor{green}{(\sqrt)}$ & $\textcolor{green}{(\sqrt)}$ & $\textcolor{red}{(\times)}$ & 0.901
        \\
        \hline
        & $\textcolor{green}{(\sqrt)}$ & $\textcolor{green}{(\sqrt)}$ & $\textcolor{green}{(\sqrt)}$ & \textbf{0.906}
        \end{tabular}
    }}
\end{table*}

\section{Conclusion}
In this work, we invoke the inherent frailty of regular convolutional neural networks in capturing long-range contextual dependencies.
Specifically, the presented methodology is a novel contextual attention network which exploits the Transformer module along with the CNN encoder in order to concomitantly combine local and global representations for a further performance boost.
The results presented in this paper validate that our proposal achieves substantial improvement over many architectures in
semantic segmentation tasks.


\bibliographystyle{splncs04}
\bibliography{Ref}

\begin{thebibliography}{10}
\providecommand{\url}[1]{\texttt{#1}}
\providecommand{\urlprefix}{URL }
\providecommand{\doi}[1]{https://doi.org/#1}

\bibitem{asadi2020multi}
Asadi-Aghbolaghi, M., Azad, R., Fathy, M., Escalera, S.: Multi-level context
  gating of embedded collective knowledge for medical image segmentation. arXiv
  preprint arXiv:2003.05056  (2020)

\bibitem{azad2019bi}
Azad, R., Asadi-Aghbolaghi, M., Fathy, M., Escalera, S.: Bi-directional
  convlstm u-net with densely connected convolutions. In: 2019 IEEE/CVF
  International Conference on Computer Vision Workshop (ICCVW). pp. 406--415
  (2019). \doi{10.1109/ICCVW.2019.00052}

\bibitem{azad2021deep}
Azad, R., Bozorgpour, A., Asadi-Aghbolaghi, M., Merhof, D., Escalera, S.: Deep
  frequency re-calibration u-net for medical image segmentation. In:
  Proceedings of the IEEE/CVF International Conference on Computer Vision. pp.
  3274--3283 (2021)

\bibitem{azad2021smu}
Azad, R., Khosravi, N., Merhof, D.: Smu-net: Style matching u-net for brain
  tumor segmentation with missing modalities  (2021)

\bibitem{ba2016layer}
Ba, J.L., Kiros, J.R., Hinton, G.E.: Layer normalization. arXiv preprint
  arXiv:1607.06450  (2016)

\bibitem{bozorgpour2021multi}
Bozorgpour, A., Azad, R., Showkatian, E., Sulaiman, A.: Multi-scale regional
  attention deeplab3+: Multiple myeloma plasma cells segmentation in
  microscopic images. arXiv preprint arXiv:2105.06238  (2021)

\bibitem{cai2020dense}
Cai, S., Tian, Y., Lui, H., Zeng, H., Wu, Y., Chen, G.: Dense-unet: a novel
  multiphoton in vivo cellular image segmentation model based on a
  convolutional neural network. Quantitative imaging in medicine and surgery
  \textbf{10}(6), ~1275 (2020)

\bibitem{cai2020ma}
Cai, Y., Wang, Y.: Ma-unet: An improved version of unet based on multi-scale
  and attention mechanism for medical image segmentation. arXiv preprint
  arXiv:2012.10952  (2020)

\bibitem{chen2021crossvit}
Chen, C.F.R., Fan, Q., Panda, R.: Crossvit: Cross-attention multi-scale vision
  transformer for image classification. In: Proceedings of the IEEE/CVF
  International Conference on Computer Vision. pp. 357--366 (2021)

\bibitem{chen2021transunet}
Chen, J., Lu, Y., Yu, Q., Luo, X., Adeli, E., Wang, Y., Lu, L., Yuille, A.L.,
  Zhou, Y.: Transunet: Transformers make strong encoders for medical image
  segmentation. arXiv preprint arXiv:2102.04306  (2021)

\bibitem{chen2017deeplab}
Chen, L.C., Papandreou, G., Kokkinos, I., Murphy, K., Yuille, A.L.: Deeplab:
  Semantic image segmentation with deep convolutional nets, atrous convolution,
  and fully connected crfs. IEEE transactions on pattern analysis and machine
  intelligence  \textbf{40}(4),  834--848 (2017)

\bibitem{codella2019skin}
Codella, N., Rotemberg, V., Tschandl, P., Celebi, M.E., Dusza, S., Gutman, D.,
  Helba, B., Kalloo, A., Liopyris, K., Marchetti, M., et~al.: Skin lesion
  analysis toward melanoma detection 2018: A challenge hosted by the
  international skin imaging collaboration (isic). arXiv preprint
  arXiv:1902.03368  (2019)

\bibitem{codella2018skin}
Codella, N.C., Gutman, D., Celebi, M.E., Helba, B., Marchetti, M.A., Dusza,
  S.W., Kalloo, A., Liopyris, K., Mishra, N., Kittler, H., et~al.: Skin lesion
  analysis toward melanoma detection: A challenge at the 2017 international
  symposium on biomedical imaging (isbi), hosted by the international skin
  imaging collaboration (isic). In: 2018 IEEE 15th International Symposium on
  Biomedical Imaging (ISBI 2018). pp. 168--172. IEEE (2018)

\bibitem{dosovitskiy2020image}
Dosovitskiy, A., Beyer, L., Kolesnikov, A., Weissenborn, D., Zhai, X.,
  Unterthiner, T., Dehghani, M., Minderer, M., Heigold, G., Gelly, S., et~al.:
  An image is worth 16x16 words: Transformers for image recognition at scale.
  arXiv preprint arXiv:2010.11929  (2020)

\bibitem{feyjie2020semi}
Feyjie, A.R., Azad, R., Pedersoli, M., Kauffman, C., Ayed, I.B., Dolz, J.:
  Semi-supervised few-shot learning for medical image segmentation. arXiv
  preprint arXiv:2003.08462  (2020)

\bibitem{gupta2018pcseg}
Gupta, A., Mallick, P., Sharma, O., Gupta, R., Duggal, R.: Pcseg: Color model
  driven probabilistic multiphase level set based tool for plasma cell
  segmentation in multiple myeloma. PloS one  \textbf{13}(12),  e0207908 (2018)

\bibitem{hatamizadeh2022unetr}
Hatamizadeh, A., Tang, Y., Nath, V., Yang, D., Myronenko, A., Landman, B.,
  Roth, H.R., Xu, D.: Unetr: Transformers for 3d medical image segmentation.
  In: Proceedings of the IEEE/CVF Winter Conference on Applications of Computer
  Vision. pp. 574--584 (2022)

\bibitem{he2016deep}
He, K., Zhang, X., Ren, S., Sun, J.: Deep residual learning for image
  recognition. In: Proceedings of the IEEE conference on computer vision and
  pattern recognition. pp. 770--778 (2016)

\bibitem{hu2018squeeze}
Hu, J., Shen, L., Sun, G.: Squeeze-and-excitation networks. In: Proceedings of
  the IEEE conference on computer vision and pattern recognition. pp.
  7132--7141 (2018)

\bibitem{huang2020unet}
Huang, H., Lin, L., Tong, R., Hu, H., Zhang, Q., Iwamoto, Y., Han, X., Chen,
  Y.W., Wu, J.: Unet 3+: A full-scale connected unet for medical image
  segmentation. In: ICASSP 2020-2020 IEEE International Conference on
  Acoustics, Speech and Signal Processing (ICASSP). pp. 1055--1059. IEEE (2020)

\bibitem{lei2020skin}
Lei, B., Xia, Z., Jiang, F., Jiang, X., Ge, Z., Xu, Y., Qin, J., Chen, S.,
  Wang, T., Wang, S.: Skin lesion segmentation via generative adversarial
  networks with dual discriminators. Medical Image Analysis  \textbf{64},
  101716 (2020)

\bibitem{li2021accurate}
Li, M., Lian, F., Wang, C., Guo, S.: Accurate pancreas segmentation using
  multi-level pyramidal pooling residual u-net with adversarial mechanism. BMC
  Medical Imaging  \textbf{21}(1), ~1--8 (2021)

\bibitem{mendoncca2013ph}
Mendon{\c{c}}a, T., Ferreira, P.M., Marques, J.S., Marcal, A.R., Rozeira, J.:
  Ph 2-a dermoscopic image database for research and benchmarking. In: 2013
  35th annual international conference of the IEEE engineering in medicine and
  biology society (EMBC). pp. 5437--5440. IEEE (2013)

\bibitem{oktay2018attention}
Oktay, O., Schlemper, J., Folgoc, L.L., Lee, M., Heinrich, M., Misawa, K.,
  Mori, K., McDonagh, S., Hammerla, N.Y., Kainz, B., et~al.: Attention u-net:
  Learning where to look for the pancreas. arXiv preprint arXiv:1804.03999
  (2018)

\bibitem{ronneberger2015u}
Ronneberger, O., Fischer, P., Brox, T.: U-net: Convolutional networks for
  biomedical image segmentation. In: International Conference on Medical image
  computing and computer-assisted intervention. pp. 234--241. Springer (2015)

\bibitem{sinha2020multi}
Sinha, A., Dolz, J.: Multi-scale self-guided attention for medical image
  segmentation. IEEE journal of biomedical and health informatics
  \textbf{25}(1),  121--130 (2020)

\bibitem{valanarasu2021medical}
Valanarasu, J.M.J., Oza, P., Hacihaliloglu, I., Patel, V.M.: Medical
  transformer: Gated axial-attention for medical image segmentation. In:
  International Conference on Medical Image Computing and Computer-Assisted
  Intervention. pp. 36--46. Springer (2021)

\bibitem{valanarasu2020kiu}
Valanarasu, J.M.J., Sindagi, V.A., Hacihaliloglu, I., Patel, V.M.: Kiu-net:
  Towards accurate segmentation of biomedical images using over-complete
  representations. In: International Conference on Medical Image Computing and
  Computer-Assisted Intervention. pp. 363--373. Springer (2020)

\bibitem{wang2018non}
Wang, X., Girshick, R., Gupta, A., He, K.: Non-local neural networks. In:
  Proceedings of the IEEE conference on computer vision and pattern
  recognition. pp. 7794--7803 (2018)

\bibitem{wu2022fat}
Wu, H., Chen, S., Chen, G., Wang, W., Lei, B., Wen, Z.: Fat-net: Feature
  adaptive transformers for automated skin lesion segmentation. Medical Image
  Analysis  \textbf{76},  102327 (2022)

\bibitem{zheng2021rethinking}
Zheng, S., Lu, J., Zhao, H., Zhu, X., Luo, Z., Wang, Y., Fu, Y., Feng, J.,
  Xiang, T., Torr, P.H., et~al.: Rethinking semantic segmentation from a
  sequence-to-sequence perspective with transformers. In: Proceedings of the
  IEEE/CVF Conference on Computer Vision and Pattern Recognition. pp.
  6881--6890 (2021)

\end{thebibliography}

\newpage
\section{Appendix}
In this part, we intend to provide some additional details regarding our approach, which allow a deeper understanding into our experiments.
\begin{figure}
\label{fig:fig1}
	\centering
	\begin{tabular}{cc}
		\includegraphics[width=1\textwidth]{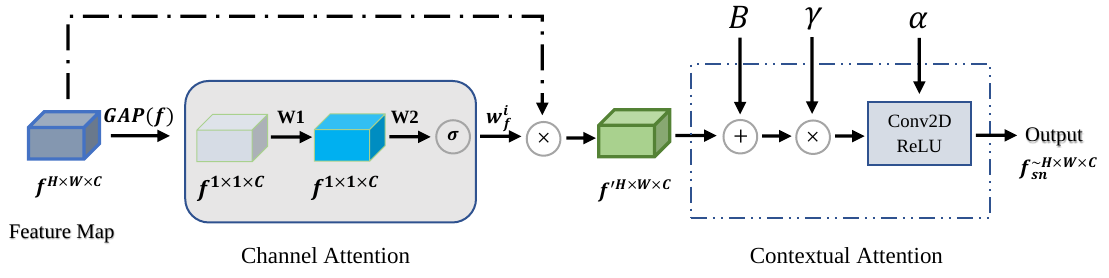}&
	\end{tabular}
	\caption{Perceptual visualization of the proposed Contextual Attention module. The proposed structure applies a channel-wise normalization along with the boundary ($B$) representation to recalibrate the feature space and then uses RIC($\gamma$) and ICR($\alpha$) features to incorporate the spatial contextual information inside the feature set.}
\end{figure}

\begin{figure}
\label{fig:fig1}
	\centering
	\begin{tabular}{cc}
		\includegraphics[width=0.93\textwidth]{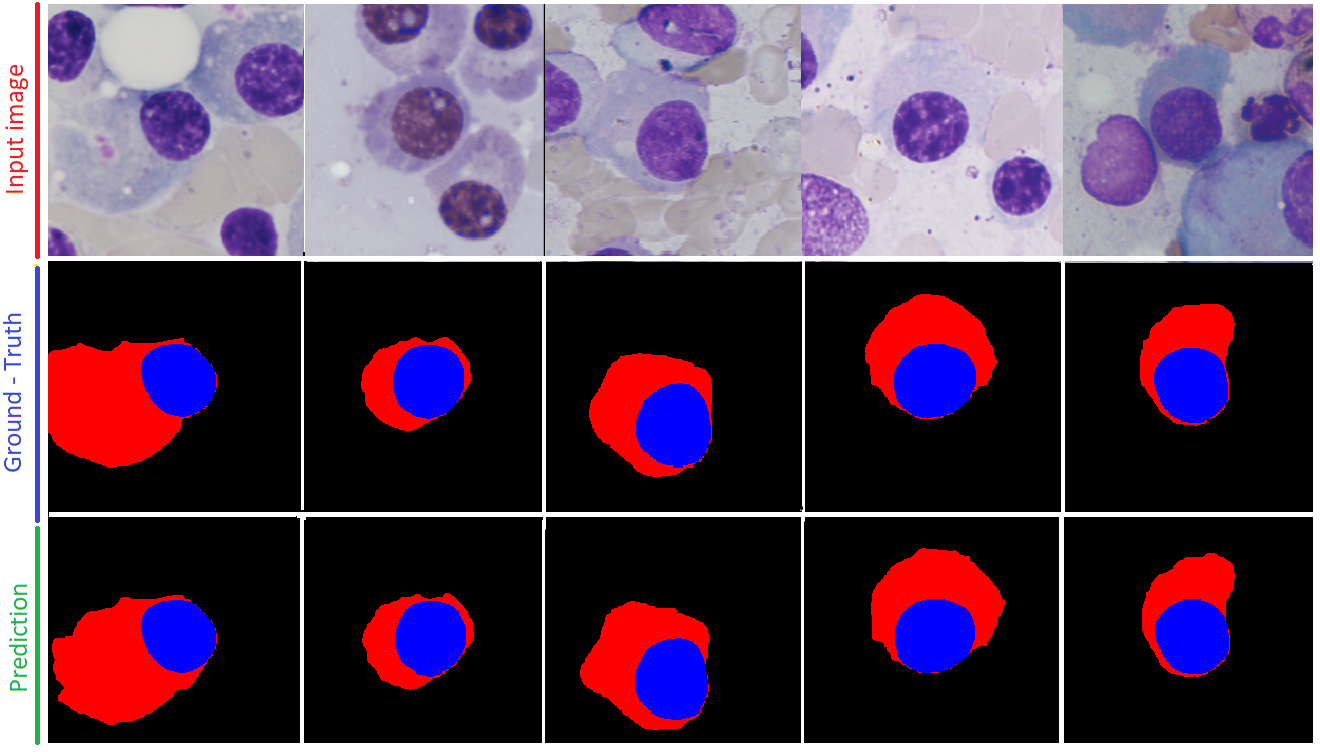}&
	\end{tabular}
	\caption{More results of the proposed method for multiple Mylomia segmentation on the SegPC2021 dataset. The first row shows the input image, the second row indicates the ground truth for each image and the third row shows the prediction of the network.}
\end{figure}

\begin{figure}[ht]
\label{fig:fig1}
	\centering
	\begin{tabular}{cc}
		\includegraphics[width=1\textwidth]{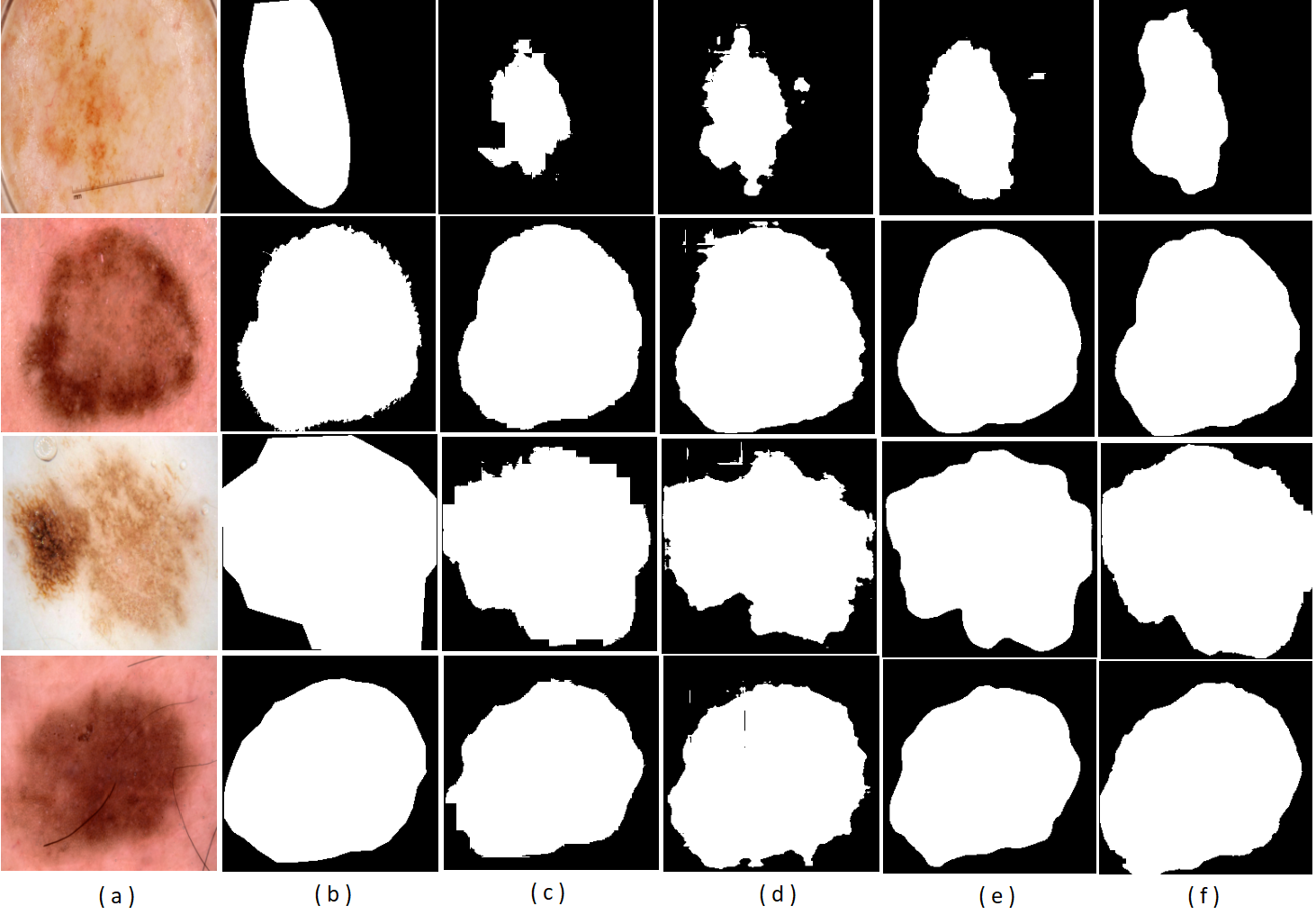}&
	\end{tabular}
	\caption{Visual comparisons of different methods for skin lesion segmentation task. (a) Input images. (b) Ground truth. (c) U-Net \cite{ronneberger2015u}. (d) Gated Axial-Attention paper \cite{valanarasu2021medical}. (e) Proposed method without a contextual attention module and (f) Proposed method.}
\end{figure}

\end{document}